\documentclass[12pt]{article}
\usepackage[margin=0.6in]{geometry}

\usepackage{notes2bib}
\usepackage{graphicx}
\usepackage[style=phys]{biblatex}
\bibliography{bibliography}
\usepackage{amsmath,amsfonts,amssymb}
\usepackage{authblk}

\usepackage{placeins}
\usepackage{xparse}
\DeclareMathAlphabet{\mathcal}{OMS}{cmsy}{m}{n}
\newcommand{\uv}[1]{\mathbf{\hat{#1}}}
\renewcommand{\v}[1]{\mathbf{#1}}
\newcommand{\vs}[1]{\boldsymbol{#1}}
\newcommand{\pd}[2]{\frac{\partial #1}{\partial #2}}
\newcommand{\s}[1]{\mathrm{#1}}
\renewcommand{\mod}[1]{|#1|}
\NewDocumentCommand{\evalat}{sO{\big}mm}{%
	\IfBooleanTF{#1}
	{\mleft. #3 \mright|_{#4}}
	{#3#2|_{#4}}%
}
\newcommand{\appropto}{\mathrel{\vcenter{
  \offinterlineskip\halign{\hfil$##$\cr
    \propto\cr\noalign{\kern2pt}\sim\cr\noalign{\kern-2pt}}}}}
\DeclareMathAlphabet\mathbfcal{OMS}{cmsy}{b}{n}
    
\author[1]{\small{Thomas J. Sturges}}
\author[2]{Taavi Rep{\"a}n}
\author[4]{Charles A. Downing}
\author[2,3]{Carsten Rockstuhl}
\author[1]{Magdalena Stobi{\'n}ska}

\affil[1]{Institute of Theoretical Physics, University of Warsaw, ul. Pasteura 5, 02-093, Warsaw, Poland}
\affil[2]{Institute of Nanotechnology, Karlsruhe Institute of Technology, 76344 Eggenstein-Leopoldshafen, Germany}
\affil[3]{Institute of Theoretical Solid State Physics, Karlsruhe Institute of Technology, 76131 Karlsruhe, Germany}
\affil[4]{Departamento de F\'{\i}sica de la Materia Condensada, CSIC-Universidad de Zaragoza, Zaragoza E-50009, Spain}

\begin{document}

\title{\textbf{Extreme renormalisations of dimer eigenmodes \\by strong light-matter coupling}}

\maketitle

\begin{abstract}
We explore by theoretical means an extreme renormalisation of the eigenmodes of a dimer of dipolar meta-atoms due to strong light-matter interactions. Firstly, by tuning the height of an enclosing photonic cavity, we can lower the energy level of the symmetric `bright' mode underneath that of the anti-symmetric `dark' mode. This is possible due to the polaritonic nature of the symmetric mode, that shares simultaneously its excitation with the cavity and the dimer. For a heterogeneous dimer, we show that the polariton modes can be smoothly tuned from symmetric to anti-symmetric, resulting in a variable mode localisation from extended throughout the cavity to concentrated around the vicinity of the dimer. In addition, we reveal a critical point where one of the meta-atoms becomes `shrouded', with no response to a driving electric field, and thus the field re-radiated by the dimer is only that of the other meta-atom. We provide an exact analytical description of the system from first principles, as well as full-wave electromagnetic simulations that show a strong quantitative agreement with the analytical model. Our description is relevant for any physical dimer where dipolar interactions are the dominant mechanism.
\end{abstract}

\vspace{2pc}

%
\maketitle

%

\section{Introduction}

The electromagnetic environment can alter the physical and chemical properties of atoms, molecules and other emitters. The ability to tailor the states of emitters through strong light-matter interactions could lead to applications in fields as diverse as chemistry, sensing and photonics \cite{Dovzhenko2018}. Indeed, there have already been concrete demonstrations such as reduced energy losses in photovoltaics \cite{Nikolis2019}, increased conductivity of organic semiconductors \cite{Orgiu2015} and enhanced non-radiative energy transfer between spatially separated molecules \cite{Zhong2017}. It has even led to the development of new research fields such as `polaritonic chemistry' \cite{Hertzog2019, Herrera2016, Kowalewski2016, Feist2018, Felipe2020, Flick2017}. This paradigm can also be exploited in metamaterials. Traditionally, the optical properties of metamaterials are realised by virtue of designing the underlying geometry of optical components \cite{Liu2011}. However, instead of ingenious geometrical designs, it has been demonstrated that one can also induce non-trivial, even topological, changes within an optical medium simply by tuning the light-matter interaction strength \cite{Downing2019, Mann2018}.

In this article, we theoretically demonstrate extreme renormalisations of the eigenmodes of a dimer of dipolar meta-atoms by strong light-matter interactions when placed inside a cavity. We consider dipolar meta-atoms as a prototypical system with the purpose to keep our study as general as possible. Such dipolar meta-atoms can accurately model many artificial systems, such as plasmonic nanoparticles \cite{Kelly2003, Fruhnert2015}, microwave helical resonators \cite{Mann2018}, magnonic microspheres \cite{Pirmoradian2018}, as well as Rydberg \cite{Leseleuc2019, Browaeys2016} and cold atoms \cite{Perczel2017}. They are the simplest building block of more complex systems, yet their interaction with light already presents some fascinating properties. We gain an understanding of such a system with a three-mode analytical model of two point dipoles interacting with the fundamental photonic mode of a cubic metallic cavity. 

Our general analytical model is enriched by a deeper numerical analysis of a specific physical system: electromagnetic simulations of a dimer of metallic nanospheroids. The numerics also confirm a strong quantitative agreement between the analytical model and full-wave electromagnetic simulations. We choose to focus on nanoparticle dimers as they have already enjoyed a wealth of investigation. The nanoparticle dimer response hybridises into a `bright' bonding and `dark' anti-bonding mode \cite{Nordlander2004}, whose frequencies depend on the interparticle seperation \cite{Rechberger2003, Cunningham2011}, with a polarisation dependence due to the anisotropic configuration \cite{Tamaru2002}. Interesting effects have been predicted and observed, such as inter-particle seperation dependent oscillations in the spectral width \cite{Olk2008} and between super- and sub-radiant damping \cite{Dahmen2007}, Fano profiles in the near-field coupling \cite{Bachelier2008}, and non-linear four-wave mixing \cite{Danckwerts2007}. In addition, quantum treatments have studied effects such as tunnelling and screening \cite{Zuloaga2009} for closely spaced dimers, as well as the Landau damping that becomes important for very small nanoparticles \cite{Brandstetter2015}. We also note that the effect of the photonic environment on the energy splitting between the dark and bright mode has demonstrated to be measurable with current technologies, despite the effect on single particle resonances being very small \cite{Downing2017}.

We show that the energy of the `bright' symmetric mode of the dimer can be shifted below that of the `dark' anti-symmetric mode, which could have profound consequences for the optical properties of the system, as it removes a non-radiative decay channel of the bright mode. Such a mechanism could be used, for example, to improve the photoluminescence of photonic devices \cite{Shahnazaryan2019}. There is another interesting effect when we consider a heterogeneous dimer, comprised of two meta-atoms with inequivalent resonant frequencies. In this case, there is an anti-crossing not only between the cavity mode and the symmetric mode, but also a second anti-crossing between the symmetric mode and the anti-symmetric mode. This results in two polaritonic modes which can be smoothly transformed from symmetric to anti-symmetric. This also manifests in a tunable mode width, from extended throughout the cavity to concentrated in the immediate vicinity of the dimer. Such a property could be used to transduce bulk changes into localised resonance shifts, and exploited in sensor applications. Moreover, there is a critical `shrouded point' where the polariton mode is neither symmetric nor anti-symmetric. Rather, the field re-radiated by the dimer is near-identical to that of a single meta-atom, as if the other meta-atom was not there . This evokes ideas of metamaterial cloaking such as those investigated by Pendry \textit{et al.} \cite{Pendry2006}, although the mechanism here is quite distinct. Whilst the renormalisation of dimer energy levels due to a photonic environment has previously been studied \cite{Downing2017}, the novelty in our work is the demonstration that the bright and dark energy levels can be completely inverted in the strong coupling regime, as well as the existence of a novel `shrouded point'.

The rest of this article is organised as follows. Section \ref{sec:analytical_model} outlines the key concepts, expected qualitative results, and introduces the analytical model. Section \ref{sec:simulations} introduces the system investigated numerically. A dimer of metallic nanospheroids is used as a case study for deeper analysis, whereas the analytical results can be generalised to many other systems. Section \ref{sec:homo} presents the results pertaining to a homogeneous dimer such as the inversion of the energies of the symmetric and anti-symmetric eigenmodes; whereas section \ref{sec:hetero} discusses the heterogeneous dimer properties, such as polariton modes with a tunable weighting of symmetric and anti-symmetric components. We provide an overview and outlook in section \ref{sec:discussion}. The Supplementary Material \cite{supp} contains a detailed derivation of our Hamiltonian from first principles, further discussions on the validity of our assumptions, and examines the effects of gain and loss.

\section{Analytical model}\label{sec:analytical_model}

\begin{figure*}[t]
	\centering
	\includegraphics[width=\textwidth]{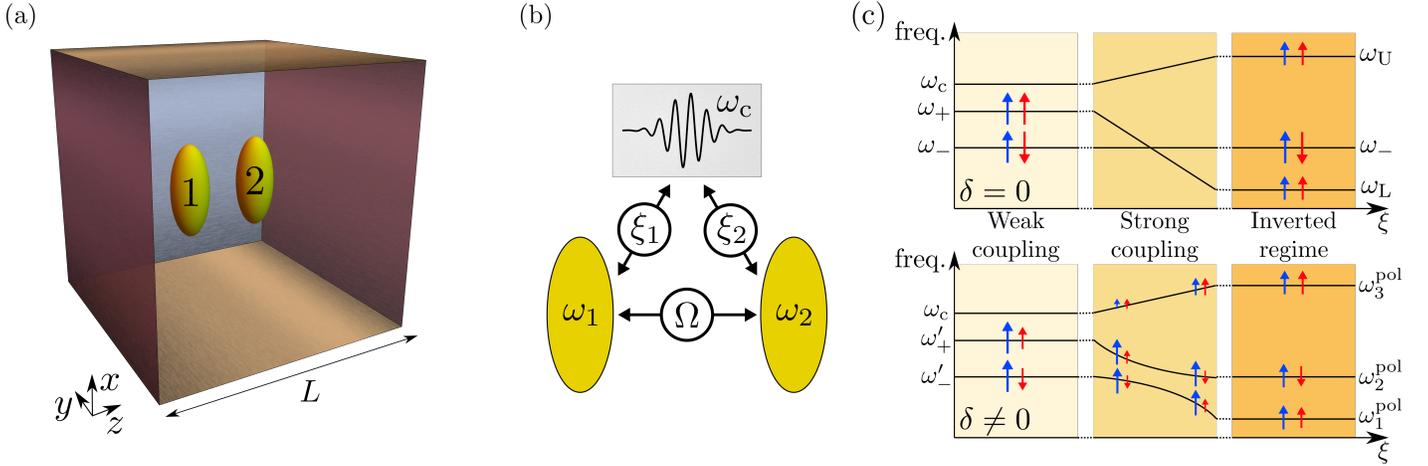}
	\caption[]{\textbf{Schematic of the model.} (a) A dimer of dipolar meta-atoms is embedded at the centre of a cubic cavity of height $L$. (b) The meta-atoms have resonant frequencies $\omega_n$, and a Coulomb interaction strength $\Omega$. The cavity mode ($\omega_\s{c}$) couples to the meta-atoms with a strength $\xi_n$. (c) Schematic of the results. Top panel: A homogeneous dimer has a symmetric ($\omega_+$) and an anti-symmetric ($\omega_-$) mode. Hybridisation with the photonic mode leads to a symmetric upper ($\omega_\s{U}$) and lower ($\omega_\s{L}$) polariton mode, whilst the `dark' anti-symmetric mode remains unchanged. In this way, the ordering of the symmetric and anti-symmetric modes can be inverted. Bottom panel: The photonic mode hybridises with both modes of a heterogeneous dimer ($\omega_\pm'$). An additional anti-crossing results in three polaritonic modes ($\omega_m^\s{pol}$) that can be smoothly tuned from symmetric to anti-symmetric.}
	\label{fig:dimer_schematic}
\end{figure*}

For simplicity, we model the photonic cavity as a cubic metal box with height $L$, see Figure \ref{fig:dimer_schematic}(a). We consider the meta-atoms as two point dipoles, separated by a distance $d$, centred in the cavity at $\v{r}_n = (L/2)(1, 1, 1 +(-1)^n d/L)$, where $n \in \{1,2\}$ labels the meta-atoms. The dipoles are associated with bosonic excitations with a resonance frequency $\omega_n$, and corresponding creation operators $b_n^\dag$. We consider small cavities such that there is no spectral overlap between the resonances of different electromagnetic eigenmodes, with the spacing between these resonances much larger than the splitting between the dimer eigenfrequencies \cite{supp}. In this way, we can consider the cavity as single mode and tune the cavity height such that the fundamental cavity mode, with frequency $\omega_c = \pi c \sqrt{2}/L$ (with $c$ the speed of light in vacuum) and creation operator $c^\dagger$, is in near-resonance with the
dipole resonant frequencies, see Figure \ref{fig:dimer_schematic}(b). In the rotating-wave approximation \cite{Hopfield1958}, and in the basis $\vs{\Psi} = (b_1,b_2,c)$, the matrix Hamiltonian reads  

\begin{eqnarray}
\label{eqn:StartingHamiltonian}
\mathcal{H}_{\vs{\Psi}} &= \left(\begin{array}{cc|c}
& & \s{i} \xi_1  \\
\multicolumn{2}{c|}{\smash{\raisebox{.5\normalbaselineskip}{$\mathcal{H}_\s{dp}$}}} & \s{i} \xi_2 \\
\hline \\[-\normalbaselineskip]
-\s{i} \xi_1 & -\s{i} \xi_2 & \omega_c 
\end{array}\right),
\\
\label{eqn:dipolar_Hamiltonian}
\mathcal{H}_\s{dp} &= \left(\begin{array}{cc}
\omega_1 & \Omega \\
\Omega & \omega_2
\end{array}\right).
\end{eqnarray}

\noindent Here, $\Omega = \sqrt{\Omega_1 \Omega_2}$ parameterises the quasistatic Coulomb interactions, where $\Omega_n = (\omega_n/2)(a_n/d)^3$, with $a_n$ a length scale that characterises the strength of dipolar excitations. A full derivation of the Hamiltonian from first principles is laid out in the Supplementary Material \bibnote[supp]{See Supplementary Material for the derivation of equation \ref{eqn:StartingHamiltonian} from first principles, an analysis of the regimes of validity, and the effects of gain and loss}. The coupling constants $\xi_n$ between the cavity mode and each meta-atom (labelled by $n$) can be analytically expressed as \citenote{supp}

\begin{equation}
    \xi_n = \left( 8\pi \frac{\Omega_n}{\omega_\s{c}}\frac{d^3}{L^3} \right)^{1/2} \cos\left( \frac{\pi d}{2 L} \right).
\end{equation}

\noindent We note that if the cavity height becomes too small than the dipole approximation breaks down. For closely space particles we thus require $L \gg d$. Under this assumption, $\xi_n$ is a monotonically decreasing function of $L$. For the parameters we will consider in this work, $\xi_n \propto 1/L$ to a good degree of accuracy. We must also be careful not to extend this model to arbitrarily large cavity heights where the single cavity mode approximation fails \citenote{supp}. Gain and loss can be easily incorporated with e.g. a Lindblad master equation approach \cite{supp}. In this case, the (Hermitian) eigenvalue spectrum still dictates the absorption peaks in the spectra, but the intensity and widths are determined by the details of driving and dissipation; and of course one must ensure that dissipation does not induce a transition from strong to weak coupling.

\section{Electromagnetic simulations}\label{sec:simulations}

The analytical model is augmented by focusing on a specific physical system of metallic nanoparticles, analysed using full-wave finite element simulations (implemented with JCMsuite software). We have checked that  strong light-matter coupling, and the results discussed herein, can be achieved by embedding the nanoparticles in a metallic cavity made of gold (Au) Fabry-P{\'e}rot mirrors with a SiO$_2$ spacer layer  \cite{Baranov2020}. However, for the sake of simplicity and easy comparison with the analytical model, we model the cavity with perfect electric conductor boundary conditions. For dipolar meta-atoms, we consider metallic ellipsoidal particles (with semi-axes $R_x=15 \mathrm{nm}$ and $R_y = R_z = 10 \mathrm{nm}$), where the permittivity is given by a simple Drude model
\begin{equation}\label{eqn:drude_model}
    \varepsilon_n(\omega) = 1 - \frac{\omega_p^2}{\omega^2 + \mathrm{i} \gamma_n}\,,
\end{equation}
with parameters $\omega_p$ and $\gamma$ chosen such that permittivity at a wavelength of 500nm is $-3.6 + 0.01\mathrm{i}$. Center-to-center distance between the particles is 35nm.
The system is excited with two dipole sources with equal dipole moments, placed at $(\pm 77.5 \mathrm{nm}, 0, 0)$. These dipoles can be either parallel or antiparallel to each other, thus allowing to excite either the bright or the dark mode of the system. We then perform frequency sweeps for each cavity height, and by using the simulated fields, calculate the total power absorbed by the particles. This allows us to identify the system resonances and plot field distributions of various modes.

\section{Homogeneous dimer}\label{sec:homo}

Let us first consider an isolated homogeneous dimer, made from meta-atoms with the same resonance frequencies $\omega_n \equiv \omega_0$ and length scales $a_n \equiv a$, and thus the same light-matter couplings $\xi_n \equiv \xi$. In this case, the dipolar Hamiltonian $\mathcal{H}_\s{dp}$ in equation \ref{eqn:dipolar_Hamiltonian} is diagonalised by the operators $\alpha_{\pm} = (1/\sqrt{2})(b_1 \pm b_2)$ with eigenfrequencies $\omega_{\pm} = \omega_0 \pm \Omega$, corresponding to a higher energy symmetric mode with aligned dipoles ($\uparrow\uparrow$) and a lower energy anti-symmetric mode with anti-aligned dipoles ($\uparrow\downarrow$). Often, these modes are referred to as the `bright' and `dark' modes due to their finite and vanishing total dipole moment respectively in the far-field. 

\begin{figure*}[ht]
	\centering
	\includegraphics[width=\textwidth]{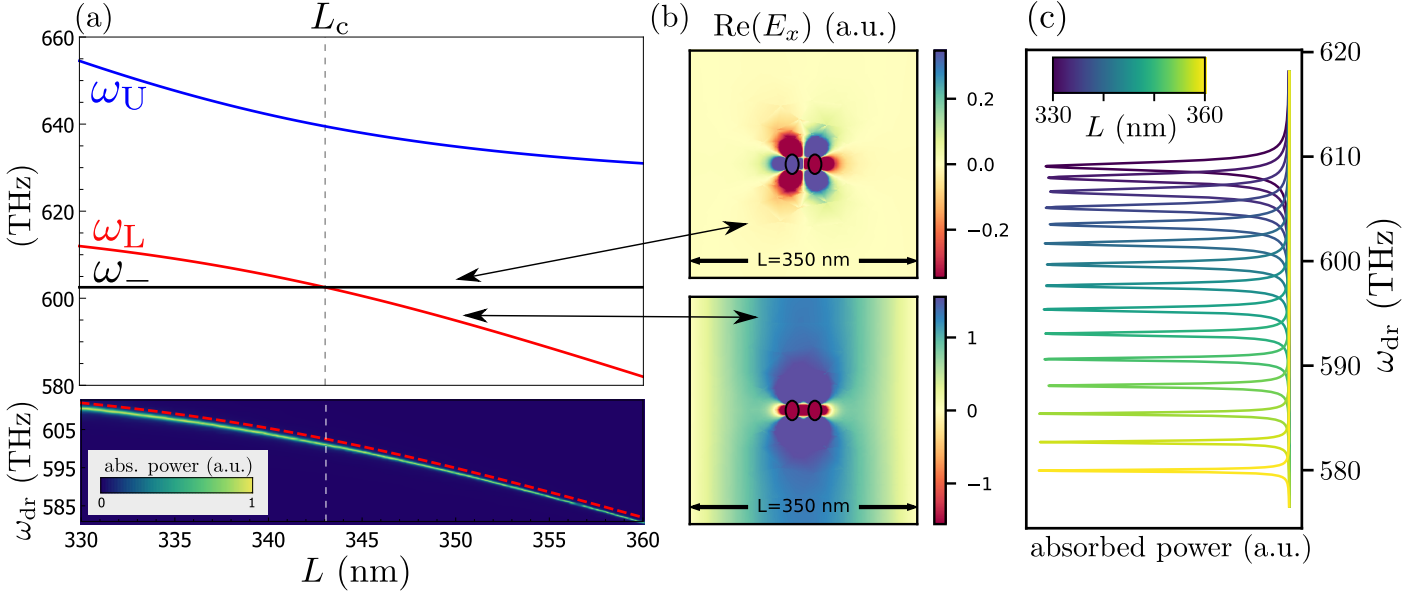}
	\caption[]{\textbf{Homogeneous dimer.} (a) Frequencies of the three polariton modes as a function of cavity height. We see the `dark' anti-symmetric mode (black), and the upper (blue) and lower (red) polariton modes. The vertical dashed line corresponds to the critical cavity height where $\omega_\s{L}$ and $\omega_-$ become degenerate. (b) The numerically calculated absorption spectra of the nanoparticles under a symmetric driving. The dashed red line corresponds to the analytically calculated lower polariton branch $\omega_\s{L}$. (c) The electric field distributions for a system driven resonantly in a symmetric (bottom) and anti-symmetric (top) configuration. (d) Frequency cuts of panel (b). In the figures $\omega_0 = 613.10\s{THz}$ and $a=11.40\s{nm}$.}
	\label{fig:polariton_dispersion}
\end{figure*}

Indeed, the anti-symmetric `dark' ($\omega_-$) mode has no coupling to the cavity mode and remains an eigenmode of the full light-matter Hamiltonian $\mathcal{H}_{\vs{\Psi}}$. This can be intuitively understood with symmetry arguments, by considering that an excitation of the symmetric cavity mode is incapable of subsequently exciting an anti-symmetric dimer mode. On the other hand, the hybridisation between the cavity and symmetric dipolar ($\omega_+$) mode results in a Rabi-splitting of magnitude $\Omega_\s{R} = \sqrt{(\omega_+-\omega_c)^2/4 + 2\xi^2}$ and a (symmetric) upper and lower polariton eigenmode with frequencies $\omega_{\s{U},\s{L}}=(1/2)(\omega_+ + \omega_c) \pm \Omega_R$.  Note that the conventional strong-coupling criteria $\Omega_\s{R} > (\gamma_1 + \gamma_2)/2$ is satisfied for the damping rates chosen in the numerical model . The Rabi splitting causes the symmetric lower polariton mode to decrease lower in energy as we increase the cavity height, see Figure \ref{fig:polariton_dispersion}(a). In fact there is a critical cavity height $L_\s{c}$ where the symmetric (lower polariton) and anti-symmetric (dark) mode become degenerate. This occurs when the Rabi frequency equals a certain detuning $\Delta = (1/2)(\omega_+ + \omega_c) - \omega_-$, namely the detuning between the dark mode and the average frequency of the bright and photon mode. For larger cavity heights, $L > L_c$, the energies of the symmetric and anti-symmetric mode become inverted.

Figure \ref{fig:polariton_dispersion}(b) overlays the analytical model and the full electromagnetic simulations. Note that there are no fitting parameters here. Rather we extract $\omega_0$ and $\omega_+$ from the resonance peaks of simulations for a single particle and dimer in free space respectively, which via the expression for $\Omega$ also tells us the value of $a$. The colour scale corresponds to the absorbed power in the metallic nanospheroids as a function of the driving frequency $\omega_\s{dr}$ of symmetrically driven point dipoles which are used to excite the system. We see that the peaks of these data follow the same trend as the (lower polariton) symmetric mode $\omega_\s{L}$. There is no absorption peak at the energy of the dark mode as this anti-symmetric mode is not excited by the symmetric driving. Cuts at fixed cavity height are displayed in Figure \ref{fig:polariton_dispersion}(d) showing in more detail how the intensity spectrum shifts with cavity height. Such data is likely the most convenient to compare to possible experimental implementations. In addition, Figure \ref{fig:polariton_dispersion}(c) shows the electric field distribution $\s{Re}(E_y)$ for a system driven resonantly in a symmetric (bottom panel) and anti-symmetric (top panel) configuration. We clearly see the hybrid light-matter nature of the symmetric mode, with the field intensity pattern extending across the whole cavity; whereas the anti-symmetric mode has a field pattern that is near-identical to its free-space counterpart, and concentrated in the vicinity of the dimer. 

\FloatBarrier

\section{Heterogeneous dimer}\label{sec:hetero}

There is no completely `dark' mode of a heterogeneous dimer, where $\omega_1 \neq \omega_2$ in equation \ref{eqn:dipolar_Hamiltonian}. In this case, the cavity mode hybridises with both dimer eigenmodes, leading to three polariton bands with an additional anti-crossing and some new physics to consider. To start our investigation, it is enlightening to rotate the full Hamiltonian $\mathcal{H}_{\vs{\Psi}}$ into the basis of symmetric and anti-symmetric operators. These operators $\alpha_\pm = (1/\sqrt{2})(b_1 \pm b_2)$ are the very same which diagonalised the homogeneous dipolar Hamiltonian. Thus in the basis $\vs{\alpha} = (\alpha_+, \alpha_-, c)$ the Hamiltonian reads

\begin{equation}
\label{eqn:HamiltonianDarkBright}
\mathcal{H}_{\vs{\alpha}} = \left(\begin{array}{ccc}
\omega_+ & \delta & \s{i} \xi_+ \\
\delta & \omega_- & \s{i} \xi_- \\
-\s{i} \xi_+ & -\s{i} \xi_- & \omega_c
\end{array}\right),
\end{equation}

\noindent where we now define $\omega_0 = (1/2)(\omega_1+\omega_2)$ as the average of the resonance frequencies of the two meta-atoms, and $\delta=(1/2)(\omega_1 - \omega_2)$ is the deviation from the average. We see that the light-matter coupling $\xi_\pm=(1/\sqrt{2})(\xi_1 \pm \xi_2)$ is large for the symmetric mode and small for the anti-symmetric mode. In this sense the symmetric mode is `brighter' whilst the other is `darker'. 

\begin{figure*}[t]
	\centering
	\includegraphics[width=\textwidth]{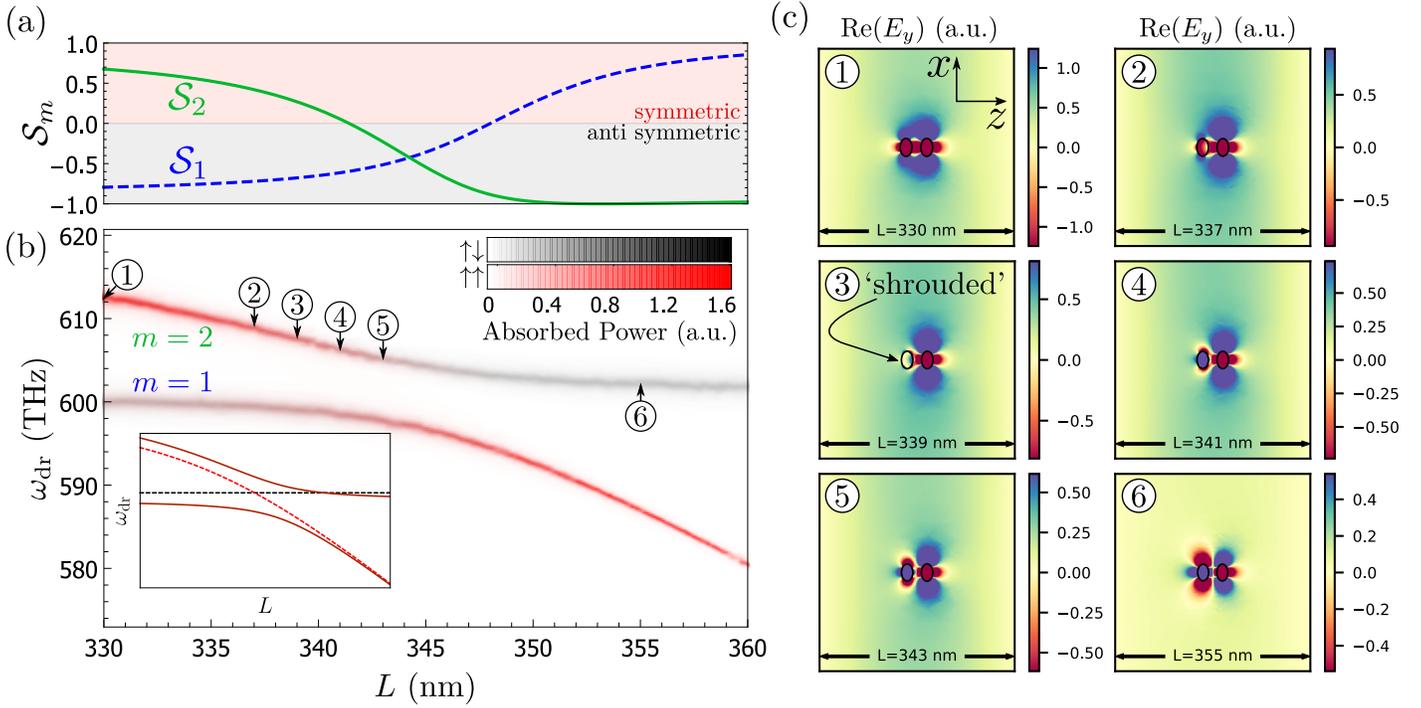}
	\caption[]{\textbf{Heterogeneous dimer.} (a) How symmetric ($0<\mathcal{S}\leq 1$) or anti-symmetric ($-1 \leq \mathcal{S} <0$) the polariton mode is. (b) Overlaid absorption spectra for the system driven symmetrically (red to transparent colour scale) and anti-symmetrically (black to transparent colour scale). Inset: analytical frequency curves (solid lines), and those of the corresponding homogeneous system for reference (dashed lines). (c) Electric field distributions for a symmetrically driven system at resonance. Black ellipses outline the nanospheroid The 3rd panel shows the `shrouded point'. In the figures $\omega_1=606.06\s{THz}$, $\omega_2=618.05\s{THz}$, $a_1=11.84\s{nm}$ and $a_2=11.81\s{nm}$.}
	\label{fig:heterogeneous_dimer}
\end{figure*}

As with the homogeneous dimer, the Rabi-splitting between the photonic mode and symmetric dimer mode causes the (originally) symmetric polariton mode to decrease in energy as the cavity height increases, whilst the (originally) anti-symmetric polariton mode has a small unimportant light-matter coupling. The key qualitative difference is that the frequency difference between the single particle resonances leads to an additional anti-crossing between the symmetric and anti-symmetric dimer mode. This can be clearly seen in the inset of Figure \ref{fig:heterogeneous_dimer}(b), which shows the dark anti-symmetric mode and the symmetric lower polariton mode of the homogeneous dimer (dashed lines), as well as the polariton bands that arise from the hybridisation between these two modes for a heterogeneous dimer (solid lines). We note that the 3rd polariton mode (not shown) is near-identical to the upper polariton mode of the homogeneous dimer (blue curve in Figure \ref{fig:polariton_dispersion}(a)), as a result of the weak coupling to the anti-symmetric dimer mode.

The result of the hybridisation between the anti/symmetric modes is that the polariton eigenmodes are neither fully symmetric nor anti-symmetric, and in-fact can be smoothly tuned from symmetric to anti-symmetric. To quantify how (anti-)symmetric a mode is we introduce the measure

\begin{equation}
    \mathcal{S}_m = \frac{\mod{v_{m+}}^2 - \mod{v_{m-}}^2}{\mod{v_{m+}}^2 + \mod{v_{m-}}^2},
\end{equation}

\noindent where $\v{v}_m = \left( v_{m+}, v_{m-}, v_{m\s{c}} \right)$ is the eigenvector of the Hamiltonian $\mathcal{H}_{\vs{\alpha}}$ corresponding to the $m$-th eigenvalue (ordered in increasing energy). A value of $\mathcal{S}_m=\pm 1$ corresponds to a fully symmetric and anti-symmetric mode respectively. Figure \ref{fig:heterogeneous_dimer}(a) explicitly shows how $\mathcal{S}_m$ varies from positive to negative (and vice versa) for the 1st and 2nd polariton modes.

Figure \ref{fig:heterogeneous_dimer}(b) shows electromagnetic simulations of the response of the system to both a symmetric and an anti-symmetric driving. Here, the nanoparticles have the same dimensions, and we alter the plasma frequency to obtain inequivalent single particle resonances; see equation \ref{eqn:drude_model}. It would correspond to a dimer made from different materials \cite{Cunningham2012}. Specifically, Figure \ref{fig:heterogeneous_dimer}(b) shows the absorption under a symmetric drive (red), overlaid on the absorption under an anti-symmetric drive (black). By plotting the absorbed power in both cases we observe how the anti/symmetric makeup of the polariton eigenmode influences the response of the system depending on the symmetry of the driving. Figure \ref{fig:heterogeneous_dimer}(c) further highlights the tunable polariton modes. As the cavity height is increased, the 2nd polariton mode changes from symmetric to anti-symmetric, resulting in an electric field distribution that contracts from extended throughout the cavity to concentrated in the vicinity of the dimer. 

The 3rd panel of Figure \ref{fig:heterogeneous_dimer}(c) shows the critical cavity height $L=339\s{nm}$ when the left meta-atom becomes `shrouded'. No electric field is excited on the left nanospheroid in response to the symmetric driving field, and the field re-radiated from the dimer is that only of a single meta-atom. Which meta-atom becomes shrouded is determined by that one with a higher (respectively lower) single meta-atom resonance frequency, when driving the system at the resonance of the lower $m=1$ (respectively middle $m=2$) polariton branch. In the analytical model this corresponds to $\mathcal{S}=0$, when the polariton eigenmode has an equal weight of both the symmetric and anti-symmetric mode ($v_{2+}=-v_{2-}$ and thus is a linear superposition of $b_2$ and $c$). Interestingly this means that the cavity has renormalised the field of a dimer to resemble that of a single meta-atom. 

\FloatBarrier

\section{Discussion}\label{sec:discussion}

We have demonstrated an extreme renormalisation of the energies of an interacting dimer of meta-atoms due to strong light-matter interactions, simply by tuning the height of an enclosing photonic cavity. For a homogeneous dimer this resulted in an inversion of the ordering of the symmetric and anti-symmetric modes; a striking fundamental demonstration that one can impart non-trivial changes to the optical properties of metamaterials, without altering the underlying geometry or symmetry of the constituents. Indeed a similar mechanism has recently been proposed to significantly enhance the photoluminescence efficiency of carbon nanotubes by removing non-radiative decay channels to a dark exciton state \cite{Shahnazaryan2019}.

In the case of a heterogeneous dimer we have seen that the polariton modes can be smoothly tuned from symmetric to anti-symmetric, resulting in a tunable mode width that varies from extended throughout the cavity to concentrated around the vicinity of the dimer. There is also a critical cavity height where one of the meta-atoms becomes `shrouded'. The mode is neither symmetric nor anti-symmetric, in fact the electric field re-radiated by the dimer resembles that of a single meta-atom, almost as if one of the meta-atoms was not there.

These phenomena should be realisable in any physical dimer where the interactions between the emitters are predominantly dipolar. In fact, qualitatively similar results should be possible in any system where one can independently tune the cavity-emitter interactions and emitter-emitter interactions. As one example, we provided full electromagnetic simulations of metallic nanoparticles. These results were in strong quantitative agreement with the analytical model derived from first principles, with no fitting parameters, showing that the simple point dipole model captured the key physics.

Our results highlight the incredible tunability offered by strong light-interactions of meta-atoms; which could find use in e.g. plasmonic transducers and sensors \cite{Du2017, Li2015}. Indeed, our scheme offers opportunities for sensing devices that transduce a change in a bulk property affecting the extended cavity mode, to that of a localised resonance shift of the dimer. Moving forward, it could be interesting to investigate quantum effects such as a possible enhancement or modulation of entanglement effects, similarly to that reported for magnons \cite{Yuan2020}.

\section*{Acknowledgements}

T.S. and M.S. were supported by the Foundation for Polish Science ``First Team'' project No.\ POIR.04.04.00-00-220E/16-00 (originally: FIRST\ TEAM/2016-2/17). C.D. acknowledges support from the Juan de la Cierva program (MINECO, Spain).
T.R. and C.R. were supported by the Helmholtz program Science and Technology of Nanosystems (STN), by the Deutsche Forschungsgemeinschaft (DFG, German Research Foundation) under Germany’s Excellence Strategy via the Excellence Cluster 3D Matter Made to Order (EXC-2082/1 – 390761711), and by the Carl Zeiss Foundation through the Carl-Zeiss-Focus@HEiKA. T.R and C.R. are also grateful to the company JCMwave for their free provision of the FEM Maxwell solver JCMsuite.

\printbibliography

\clearpage

\setcounter{section}{0}
\setcounter{page}{1}

\begin{center}
    \Large{\textbf{Supplementary Material: Extreme renormalisations of dimer eigenmodes by strong light-matter coupling}}
\end{center}

\maketitle

\begin{abstract}
This document contains the Supplementary Material to the article published under the same name. We outline the derivation of the Hamiltonian in the main text from first principles. During this process we also derive the general Hamiltonian for all cavity modes (including non-resonant terms) which could be useful for future investigations. We show how the inclusion of only the fundamental mode $\ell=(0,1,1)$ is valid for the parameters considered in the main text. Finally we extend the model to include image dipoles, and show that the omission of these and non-resonant terms was valid in the main text.
\end{abstract}

\vspace{2pc}

%
\maketitle

%

\section{Dipolar Hamiltonian}

In this section we detail the derivation of the dipolar Hamiltonian. We consider the quasistatic Coulomb interactions between two dipolar meta-atoms. We consider meta-atoms composed of physical objects whose natural frequency of the charge density oscillations along the $x$-direction are tuned in resonance with the fundamental cavity mode. Whereas the natural frequencies along the other directions are detuned and not relevant to the description. In this way we need only consider the one transverse polarization along the $x$-axis. We consider a dimer of meta-atoms labelled by $n \in \{1,2\}$, whose positions are $\v{r}_n = (1/2) \left( L_x, L_y, L_z + (-1)^n d \right)$. They have masses $M_n$, natural frequencies $\omega_n$, and charges $Q_n = N_n e^2$, where $e$ is elementary charge constant. We denote the displacement of the electronic centre of mass by $h_n$, and the conjugate momentum by $\Pi_n$. Including the quasistatic Coulomb interactions the Hamiltonian of the meta-atoms reads \cite{Brandstetter2015}

\begin{equation}
H_\s{dp} = \sum_n \left[ \frac{\Pi^2_n}{2M_n} + \frac{M_n \omega_n^2}{2} h_n^2 \right] + \frac{Q_1 Q_2}{4\pi\epsilon d^3} h_1 h_2,
\end{equation}

\noindent where $\epsilon$ is the dielectric constant of the surrounding medium. Throughout this work we work in units such that $\epsilon=1$ and $\hbar=1$ but retain the symbolic representation of $\epsilon$ for completeness. We switch to the second quantisation representation by the substitutions 

\begin{equation}\label{eqn:handPi}
\begin{array}{cc}
h_n = (1/2M_n\omega_n)^{1/2}\left( b_n + b_n^\dagger \right), & 
\Pi_n = i(M_n \omega_n/2)^{1/2}\left( b_n - b_n^\dag \right),
\end{array}
\end{equation}

\noindent which gives

\begin{equation}\label{eqn:dipolar_hamiltonian}
H_\s{dp} = \sum_n \left[ \omega_n b_n^\dag b_n \right] + \Omega \left( b_1 + b_1^\dagger \right) \left( b_2 + b_2^\dagger \right),
\end{equation}

\noindent where $\Omega = \sqrt{\Omega_1 \Omega_2}$ parameterises the strength of Coulomb interactions. Here, $\Omega_n = (\omega_n/2) (a_n / d)^3$ where $a_n = \left( Q_n^2 / 4 \pi \epsilon M_n \omega_n^2 \right)^{1/3}$ is a length scale characterising the strength of the dipolar excitations.

\section{Magnetic vector potential of a box with perfect metallic walls}

Before deriving the light-matter Hamiltonian in the subsequent section, first we must deduce the magnetic vector potential in a box with perfect metallic walls. We follow the paper by K. Kakazu and Y. S. Kim \cite{Kakazu1994} but explain the steps in greater detail. After obtaining the most general expression, we will derive the specific form used in the main text after our assumptions.

\subsection{Deriving $\v{A}(\v{r},t)$ in full generality}

We consider a box with perfect metallic walls. The cavity has sides of length $L_x$, $L_y$ and $L_z$. The boundary conditions are that the tangential component of $\v{E}$ and the normal component of $\v{B}$ vanish at the cavity walls. Maxwell's equations in free space leads to the wave equation

\begin{equation}
\left( \nabla^2 - \frac{1}{c^2} \pd{}{t} \right) \left(\begin{array}{c} 
\v{E} \\ \v{B} \end{array}\right) 
= \left(\begin{array}{c} 
0 \\ 0 
\end{array}\right).
\end{equation}

\noindent Thus for both the magnetic and electric field, for each component we need to solve an equation of the form

\begin{equation}
\left( \nabla^2 - \frac{1}{c^2} \pd{}{t} \right) \psi = \psi.
\end{equation}

\noindent Using separation of variables and assuming a harmonic time dependence, the solution is of the form $\psi(x,y,z,t) = X(x)Y(y)Z(z)\s{e}^{-\s{i} \omega t}$. Thus we have $X'' + k^2 X = 0$ with general solution

\begin{equation}
X(x) = c_1 \cos(k x) + c_2 \sin(k x),
\end{equation}

\noindent and similarly for $Y$ and $Z$. Using Maxwell's equations, particularly $\nabla \times \v{E} - \pd{\v{B}}{t} = 0$, and the boundary conditions allows us to deduce the specific form of the electric field eigenmodes as

\begin{eqnarray}\label{electric_field_eigenmodes}
\v{v}_{\ell x} &=& \mathcal{N}_{\ell x} \cos(k_{\ell x} x) \sin(k_{\ell y} y) \sin(k_{\ell z} z) \uv{x}, \\
\v{v}_{\ell y} &=& \mathcal{N}_{\ell y} \sin(k_{\ell x} x) \cos(k_{\ell y} y) \sin(k_{\ell z} z) \uv{y}, \\
\v{v}_{\ell z} &=& \mathcal{N}_{\ell z} \sin(k_{\ell x} x) \sin(k_{\ell y} y) \cos(k_{\ell z} z) \uv{z},
\end{eqnarray}

\noindent where $\mathcal{N}_{\ell i} = \sqrt{8/(V(1+\delta_{\ell i}))}$ is the normalisation coefficient, $\delta_{\ell i} \equiv \delta(\ell_i)$ is the Kronecker delta of $\ell_i$, and $k_{\ell i} = \pi \ell_i / L_i$ is the `wavenumber'. Here, $i \in \{x,y,z\}$ labels the Cartesian coordinates, and we label the photonic modes with the list of integers $\ell = \{ \ell_x, \ell_y, \ell_z \}$. The normalisation coefficient arises from imposing orthonormality $\int_\s{cavity} \s{d}V \v{v}_{\ell i} \v{v}_{\ell'i'} = \delta_{ii'} \delta_{\ell \ell'}$. Note also that not all the combinations of $(l_x,l_y,l_z)$ contribute non-zero eigenvectors (e.g. $(0,0,1)$ and its permutations). These can be easily rejected.

Thus we can construct a general solution of the electric field from a linear combination of these field eigenmodes

\begin{equation}
\v{E}(\v{r},t) = \s{i}  \sum_{\ell i} \mathcal{D}_{\ell i} \v{v}_{\ell i}(\v{r}) + \s{H.c.},
\end{equation}

\noindent where $\mathcal{D}_{\ell i}$ are the expansion coefficients, and the phase factor is introduced by convention. The transversality condition (derived from $\nabla \cdot \v{E} = 0$) is

\begin{equation}
\v{k}_\ell \cdot \left( \mathcal{D}_{\ell x} \uv{x} + \mathcal{D}_{\ell y} \uv{y} + \mathcal{D}_{\ell z} \uv{z} \right) = 0.
\end{equation}

\noindent This shows us that we are over specifying the system, and in fact only need two orthogonal unit vectors to resolve the components in the expansion of $\v{E}$. Therefore let us perform a rotation to a coordinate frame where $\uv{e}_3 = \uv{k}_\ell$ and the other two unit vectors, $\uv{e}_1$ and $\uv{e}_2$, are orthogonal to $\v{k}_\ell$. First write $\v{E}$ as

\begin{equation}
\v{E}(\v{r},t) = \s{i}  \sum_{\ell} \mathbfcal{D}_\ell \cdot \v{V}_\ell + \s{H.c.},
\end{equation}

\noindent where $\vs{\mathbfcal{D}}_\ell$ and $\v{V}_\ell$ are vectors formed from the components $\mathcal{D}_{\ell i}$ and $\v{v}_{\ell i}$ in the obvious way (note that technically $\v{V}_\ell$ is a vector of vectors). Then we introduce the orthogonal `change of basis' matrix $\v{Q}_\ell$ which (because orthogonal matrices preserve the dot product when acting on both of the vectors) allows us to write $\v{E}$ as 

\begin{equation}
\v{E}(\v{r},t) =  \s{i}  \sum_\ell \mathbfcal{C}_\ell \cdot \v{U}_\ell + \s{H.c.},
\end{equation}

\noindent where $\v{U}_\ell = \v{Q}_\ell \v{V}_\ell$ and  $\mathbfcal{C}_\ell = \v{Q}_\ell \mathbfcal{D}_\ell$; their components are $\v{u}_{\ell\lambda}$ and $\mathcal{C}_{\ell\lambda}$ respectively, where $\lambda \in \{1,2,3\}$ (we will shorten the set of $\lambda$ in a moment). Going back to the summation notation we have

\begin{equation}
\v{E}(\v{r},t) =  \s{i}  \sum_{\ell \lambda} \mathcal{C}_{\ell \lambda}(t) \v{u}_{\ell \lambda}(\v{r}) + \s{H.c.} 
\end{equation}

\noindent Finally, we choose the orthogonal matrix

\begin{equation}
\v{Q}_\v{k} = 
\left(\begin{array}{c}
\leftarrow \uv{e}_1 \rightarrow \\
\leftarrow \uv{e}_2 \rightarrow \\
\leftarrow \uv{e}_3 \rightarrow 
\end{array}\right)
=
\left(\begin{array}{ccc}
\cos\theta_\ell \cos\phi_\ell & \cos\theta_\ell \sin\phi_\ell & -\sin\theta_\ell \\
-\sin\phi_\ell & \cos\phi_\ell & 0 \\
\sin\theta_\ell \cos\phi_\ell & \sin\theta_\ell \sin\phi_\ell & \cos\theta_\ell
\end{array}\right)
\end{equation}

\noindent where $\theta_\ell = \arctan\left(k_{\ell z}, \sqrt{k_{\ell x}^2 + k_{\ell y}^2}\right)$ and $\phi_\ell = \arctan(k_{\ell x},k_{\ell y})$ are the polar and azimuthal angles of $\v{k}_\ell$. The transversality requirement is now easily satisfied by setting $\mathcal{C}_{\ell 3} = 0$. Thus from now on summations over the different modes can be restricted to the set $\lambda \in \{1,2\}$. Finally we quantise the system by promoting the expansion coefficients to annihilation operators

\begin{equation}
\mathcal{C}_{\ell\lambda} \rightarrow \sqrt{\frac{ \omega_\ell}{2\epsilon_0}} c_{\ell\lambda},
\end{equation}

\noindent where $\omega_\ell = \s{c} \mod{\v{k}_\ell}$, and $c_{\ell\lambda}$ is the annihilation operator obeying $\s{d}c_{\ell \lambda}/\s{d}t = -i \omega_\ell c_{\ell \lambda}$. The normalisation coefficient is chosen so that the Hamiltonian has the simple form

\begin{equation}\label{eqn:photonic_hamiltonian}
\boxed{H_\s{ph} = \int_\s{cavity}\s{d}V\left( \mod{\v{E}}^2 + c^2\mod{\v{B}}^2 \right) = \sum_{\ell \lambda} \omega_\ell  c_{\ell \lambda}^\dag c_{\ell \lambda}},
\end{equation}

\noindent where the global energy shift of $\sum_{\ell \lambda}(1/2)$ has been dropped. Getting back to the task in hand, we now write the electric field with operators as

\begin{equation}
\v{E}(\v{r},t) =  \s{i}  \sum_{\ell \lambda} \left( \frac{\omega_\ell}{2\epsilon_0} \right)^{1/2} c_{\ell \lambda} \v{u}_{\ell\lambda} + \s{H.c.}.
\end{equation}

\noindent From this we can automatically use Maxwell's equation, $\partial \v{B} / \partial t = -\nabla \times \v{E}$, to write the magnetic field, $\v{B} = -i \nabla \times \v{E}/\omega_\ell$, as

\begin{equation}
\v{B}(\v{r},t) =  \sum_{\ell \lambda} \left( \frac{1}{2\epsilon_0 \omega_\ell} \right)^{1/2} c_{\ell \lambda} \nabla \times \v{u}_{\ell \lambda} + \s{H.c.}
\end{equation}

\noindent As this expression already contains a curl, we readily obtain the expression for the magnetic vector potential as

\begin{equation}
\v{A}(\v{r},t) =  \sum_{\ell \lambda} \left( \frac{1}{2\epsilon_0  \omega_\ell} \right)^{1/2} c_{\ell \lambda} \v{u}_{\ell \lambda} + \s{H.c.}
\end{equation}

\noindent In the expression for $\v{A}$ we have implicitly chosen the Coulomb gauge by omitting the gradient of an arbitrary function. The mode functions are given explicitly by

\begin{equation}
\v{u}_{\ell \lambda} =  
\left(\begin{array}{c}
\mathcal{N}_{\ell x} \Lambda_{\ell \lambda x} \cos(k_{\ell x} x) \sin(k_{\ell y} y) \sin(k_{\ell z} z) \\
\mathcal{N}_{\ell y} \Lambda_{\ell \lambda y} \sin(k_{\ell x} x) \cos(k_{\ell y} y) \sin(k_{\ell z} z)  \\
\mathcal{N}_{\ell z} \Lambda_{\ell \lambda z} \sin(k_{\ell x} x) \sin(k_{\ell y} y) \cos(k_{\ell z} z) 
\end{array}\right)
\end{equation}

\noindent where the prefactors due to the two photon polarisations are

\begin{equation}
\begin{array}{cc}
\vs{\Lambda}_{\ell 1} = 
\left(\begin{array}{c}
\cos\theta_\ell \cos\phi_\ell \\
\cos\theta_\ell \sin\phi_\ell \\
-\sin\theta_\ell
\end{array}\right),
&
\vs{\Lambda}_{\ell 2} = 
\left(\begin{array}{c}
-\sin\phi_\ell \\
\cos\phi_\ell \\
0
\end{array}\right).
\end{array}
\end{equation}

\subsection{The component of $\v{A}$ aligned with the dipole polarisation at the meta-atom sites}

We now derive the expression for the component of the magnetic vector potential aligned with the dipole polarisation ($x$-axis), exactly at the location of the meta-atoms $\v{r}_n = (1/2)(L_x, L_y, L_z + \tau_n d)$, where $\tau_{1,2} = \pm 1$ labels the meta-atom. By simple substitution, we can first write the x-component of the magnetic vector potential at the nanoparticle positions as

\begin{equation}
\begin{array}{l}
	A_n \equiv \v{A}(\v{r}_n,t) \cdot \uv{x} = \\ \sum_{\ell\lambda} \mathcal{N}_{\ell x} \left(\frac{1}{2\epsilon_0 \omega_\ell}\right)^{1/2} \cos\left( \frac{\pi \ell_x}{2} \right) \sin\left( \frac{\pi \ell_y}{2} \right) \sin\left( \frac{\pi \ell_z}{2}[1 + \tau_n d/L_z] \right)  \Lambda_{\ell \lambda x} c_{\ell \lambda} +\s{H.c.}
\end{array}
\end{equation}

\noindent Several of the terms in this sequence are zero, and thus it would be useful to relabel the system to remove this redundancy. This has the practical advantage that we do not have to keep track of which modes actually have any coupling to the meta-atoms and which do not. Thus we exploit the following identities

\begin{equation}
\begin{array}{c}
	\sum_{n \in \mathbb{N}} \cos\left( \frac{\pi n}{2} \right) f(n) = \sum_{n \in \mathbb{N}} (-1)^n f(2 n), \\
	\sum_{n \in \mathbb{N}} \sin\left( \frac{\pi n}{2} \right) f(n) = \sum_{n \in \mathbb{N}} (-1)^n f(2 n + 1), \\
	\sin\left[ \frac{\pi n}{2} (1 \pm x) \right] = (\pm 1)^{n+1} \sin \left[ \frac{\pi n}{2} (1+x) \right].
\end{array}
\end{equation}

\noindent These identities allows us to write the magnetic vector potential as

\begin{equation}
    \begin{array}{l}
	A_n = \sum_{\ell \in \mathcal{P}}\sum_\lambda P^n_\ell A_{\ell \lambda}, \\
	A_{\ell \lambda} =  \left(\frac{4}{\epsilon_0 \omega_\ell V(1+\delta_{\ell x})}\right)^{1/2} \sin\left[ \frac{\pi \ell_z}{2}(1 + d/L_z) \right] \Lambda_{\ell \lambda x} \left( c_{\ell \lambda} + c_{\ell \lambda}^\dagger \right),
\end{array}
\end{equation}

\noindent where the sum over $\ell$ is restricted to the set $\mathcal{P}$ where $\ell_x$ is even, $\ell_y$ is odd, and $\ell_z$ is any natural number. Here, the parity of the mode $\ell$ with polarisation $\lambda$ at nanoparticle $n$ is encoded in the quantity $P_\ell^n = (-1)^{\ell_x + \ell_y} (\tau_n)^{\ell_z + 1}$, where obviously $P_\ell^n \in \{1,-1\}$.

\section{Light matter interaction Hamiltonian}

Now that we have derived the magnetic vector potential, we can move on to obtaining the light-matter interaction Hamiltonian. Neglecting diamagnetic terms which only introduce a negligible self-energy renormalisation of the photon frequencies, the light-matter coupling in the minimal coupling formalism is 

\begin{equation}
	H_\s{int} = \sum_n \frac{Q_n}{M_n} \Pi_n A_n = \sum_{\ell \in \mathcal{P}} \sum_{n \lambda} P_\ell^n \frac{Q_n}{M_n} \Pi_n A_{\ell \lambda},
\end{equation}

\noindent which can be expressed in terms of the bosonic operators as

\begin{equation}\label{eqn:interaction_hamiltonian}
	H_\s{int} = i \sum_{\ell \in \mathcal{P}} \sum_{n \lambda} P_\ell^n \xi_{\ell \lambda}^n (b_n - b_n^\dagger)(c_{\ell \lambda} + c_{\ell \lambda}^\dagger).
\end{equation}

\noindent The light-matter coupling strength is parameterised by

\begin{equation}\label{eqn:xi_most_general}
\xi_{\ell \lambda}^n = 4 \omega_n \left( \frac{\pi}{1+\delta_{\ell x}} \frac{\Omega_n}{\omega_\ell} \frac{d^3}{V} \right)^{1/2} \sin\left[ \frac{\pi \ell_z}{2}(1+d/L_z) \right] \Lambda_{\ell \lambda x}.
\end{equation}

\subsection{Including only the fundamental cavity mode}

We tune our system such that only the fundamental cavity mode $\ell=\{0,1,1\}$ is in near-resonance with the meta-atom resonance frequency and has a significant light-matter coupling. We will justify this in later sections. In this case the interaction Hamiltonian reads

\begin{equation}\label{eqn:fund_int_ham}
H_\s{int} = i \sum_n \xi_n (b_n - b_n^\dagger)(c + c^\dagger),
\end{equation}

\noindent where

\begin{equation}
\xi_n = \left( 8\pi \frac{\Omega_n}{\omega_c} \frac{d^3}{V} \right)^{1/2} \cos\left( \frac{\pi d}{2 L_z} \right).
\end{equation}

\noindent Here, $c^\dagger$ creates a photon in the mode $\ell=\{0,1,1\}$ with polarisation $\lambda=2$ and frequency $\omega_c \equiv \omega_{\{0,1,1\}}$ (the other polarisation has no component in the $x$-direction). The photonic Hamiltonian is trivially

\begin{equation}\label{eqn:fund_ph_ham}
    H_\s{ph} = \omega_c c^\dagger c.
\end{equation}

\subsection{Significance of the other cavity modes}

To get a feel for the action and significance of the different modes, let us assume that we have (and that it is possible to have) tuned the cavity such that we can neglect all other modes apart from (an arbitrary) $\ell$ and $\lambda$, and consider a homogenous dimer ($\omega_0 \equiv \omega_1 = \omega_2$). In this case the Hamiltonian in the rotating wave approximation, in the basis $\vs{\Psi}^\dagger = (b_1^\dagger, b_2^\dagger, c^\dagger)$, is

$\mathcal{H}_\Psi = 
\left(\begin{array}{ccc}
\omega_0 & \Omega & -i P_1 \xi \\
\Omega & \omega_0 & -i P_2 \xi \\
i P_1 \xi & i P_2 \xi & \omega_c
\end{array}\right).
$

\noindent where we have defined $P_{1,2} \equiv P_\ell^{1,2}$, as well as $\xi \equiv \xi_{\ell\lambda}^1 = \xi_{\ell\lambda}^2$ and $\omega_c \equiv \omega_\ell$. Let us write this Hamiltonian in the basis $\vs{\gamma}^\dagger = (\gamma_B^\dagger, \gamma_D^\dagger, c^\dagger)$ where the `bright' and `dark' operators are $\gamma_{B,D} = (1/\sqrt{2})(P_1 b_1 \pm P_2 b_2)$. It reads

\begin{equation}
\mathcal{H}_\gamma = 
\left(\begin{array}{ccc}
\omega_B & 0 & -i \sqrt{2} \xi \\
0 & \omega_D & 0 \\
i \sqrt{2} \xi & 0 & \omega_c
\end{array}\right),
\end{equation}

\noindent where we have defined $\omega_{B,D} = \omega_0 \pm P_1 P_2 \Omega$. Interestingly, we see that the mode which couples to an individual cavity mode is not necessarily the higher energy symmetric dipolar mode; the one we typically consider as the `bright' mode in free-space. For greater clarity we can write the new true bright and dark eigenmodes (in this context) as

\begin{equation}\label{eqn:omegaBD}
\omega_{B,D} = \omega_0 \mp (-1)^{\ell_z} \Omega.
\end{equation}

\noindent We can also separate the Hamiltonian into two block diagonal pieces; the dark mode $\omega_D$, and the Hamiltonian of the photon and bright modes. In the basis $(\gamma_B^\dagger, c^\dagger)$ this reads

\begin{equation}
\mathcal{H}_\gamma = 
\left(\begin{array}{cc}
\omega_B & -i \xi_B \\
i \xi_B & \omega_c
\end{array}\right).
\end{equation}

\begin{figure*}[t]
	\centering
	\includegraphics[width=0.5\textwidth]{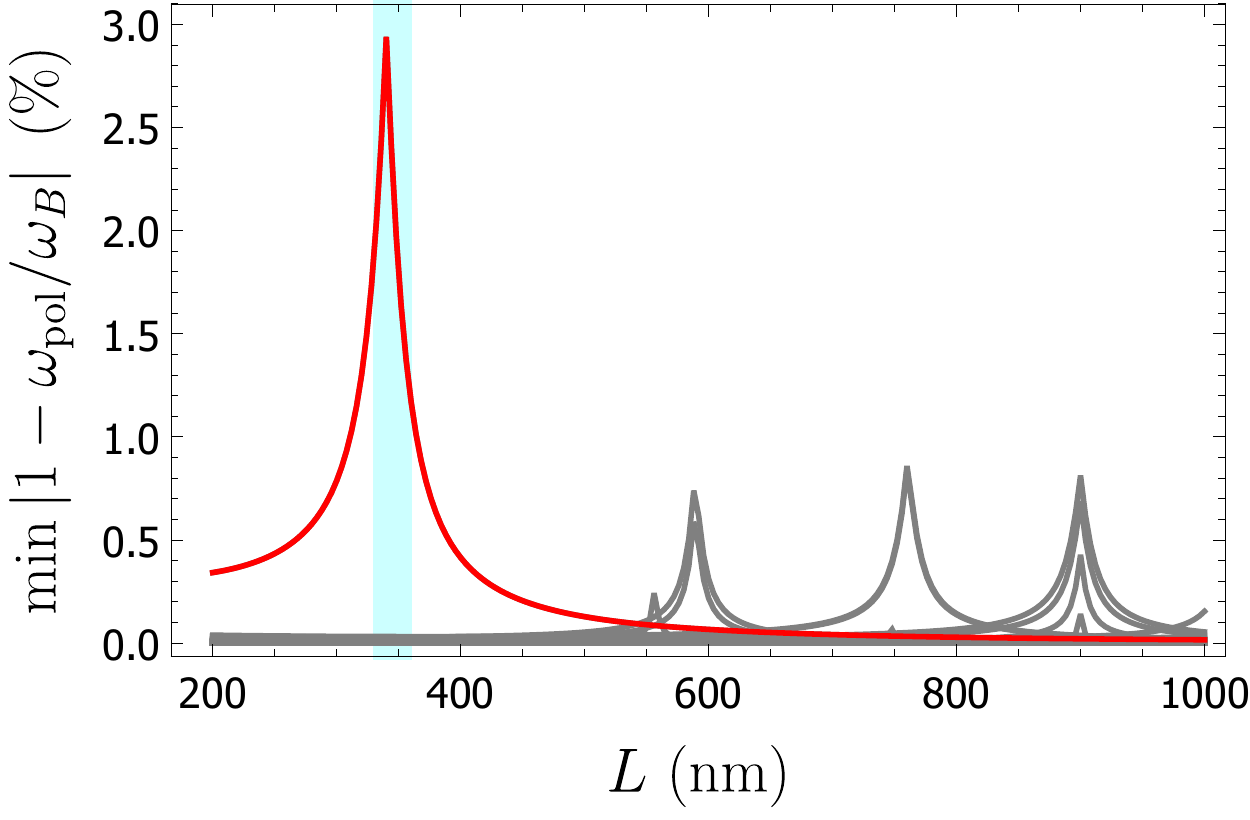}
	\caption[]{\textbf{Renormalisation strength of the various cavity modes.} We characterise the extent to which a cavity mode renormalises the bright mode of a heterogeneous dimer, when considering just that cavity mode in the model. We plot the percentage change of the polariton mode $\omega_\s{pol}$ away from the uncoupled bright-mode $\omega_B$ for $\ell_x \in \{0,2,4,6\}$, $\ell_y \in \{1,3,5,7\}$, $\ell_z \in \{0,1,2,3\}$, and both polarisations. We show the mode with $\ell=\{0,1,1\}$ and $\lambda=2$ in red, which is exactly the mode considered in the main text. All other parameters are the same as in the main text. The cyan region corresponds to the range of $L$ considered in the main text.}
	\label{fig:cavity_modes_weighting}
\end{figure*}

\noindent The polariton eigenmodes are

\begin{equation}
\omega_\s{pol} = \omega_{Bc} \pm \sqrt{\delta_{Bc}^2 + \xi_B^2},
\end{equation}

\noindent where we have defined the average frequency $\omega_{Bc} = (1/2)(\omega_B + \omega_c)$ and detuning $\delta_{Bc} = (1/2)(\omega_B - \omega_c)$. In figure \ref{fig:cavity_modes_weighting} we plot the percentage difference between this polariton mode and the uncoupled bright-mode, for several cavity modes. This gives us a measure of how significant each cavity mode is in the full general light-matter Hamiltonian of equation \ref{eqn:xi_most_general}. We see that in the range of cavity dimensions considered in the main text the mode $\ell=\{0,1,1\}$ is the only mode which causes a significant renormalisation of the bright mode.

\FloatBarrier

\section{Non-resonant terms and image dipoles}

In the main paper we have neglected non-resonant terms (such as $b_1^\dagger b_1^\dagger$) as their contributions are negligible for $\Omega,\xi \ll \omega_0$. In both the main paper and this Supplementary Material we have also neglected image dipoles. We will reintroduce both these contributions in this sections and show that they alter the spectra in insignificant ways.

The boundary conditions of the metallic box renormalise the purely dipolar electric field of the point dipole. This can be represented by image dipoles, leading to a image dipole Hamiltonian

\begin{equation}\label{eqn:image_dipoles}
H_\s{im} = -\left[ \sum_n  \frac{\Omega_n \mathcal{I}^{(0)}}{2} \left( b_n + b_n^\dagger \right) \left( b_n + b_n^\dagger \right) + \Omega \mathcal{I}^{(1)} \left( b_1 + b_1^\dagger \right) \left( b_2 + b_2^\dagger \right)\right]
\end{equation}

\noindent where

\begin{equation}
\mathcal{I}^{(s)} = \sum_{\ell \neq 0} (-1)^{\ell_y+\ell_z} \mathcal{C}\Big(\ell_x L_x, \ell_y L_y, \ell_z L_z + d \s{mod}_2 (\ell_z + s) \Big)
\end{equation}

\noindent and

\begin{equation}
\mathcal{C}(x,y,z) =  \frac{y^2+z^2-2x^2}{\left[ x^2 + y^2 + z^2 \right]^{5/2}}.
\end{equation}

\begin{figure*}[t]
	\centering
	\includegraphics[width=0.75\textwidth]{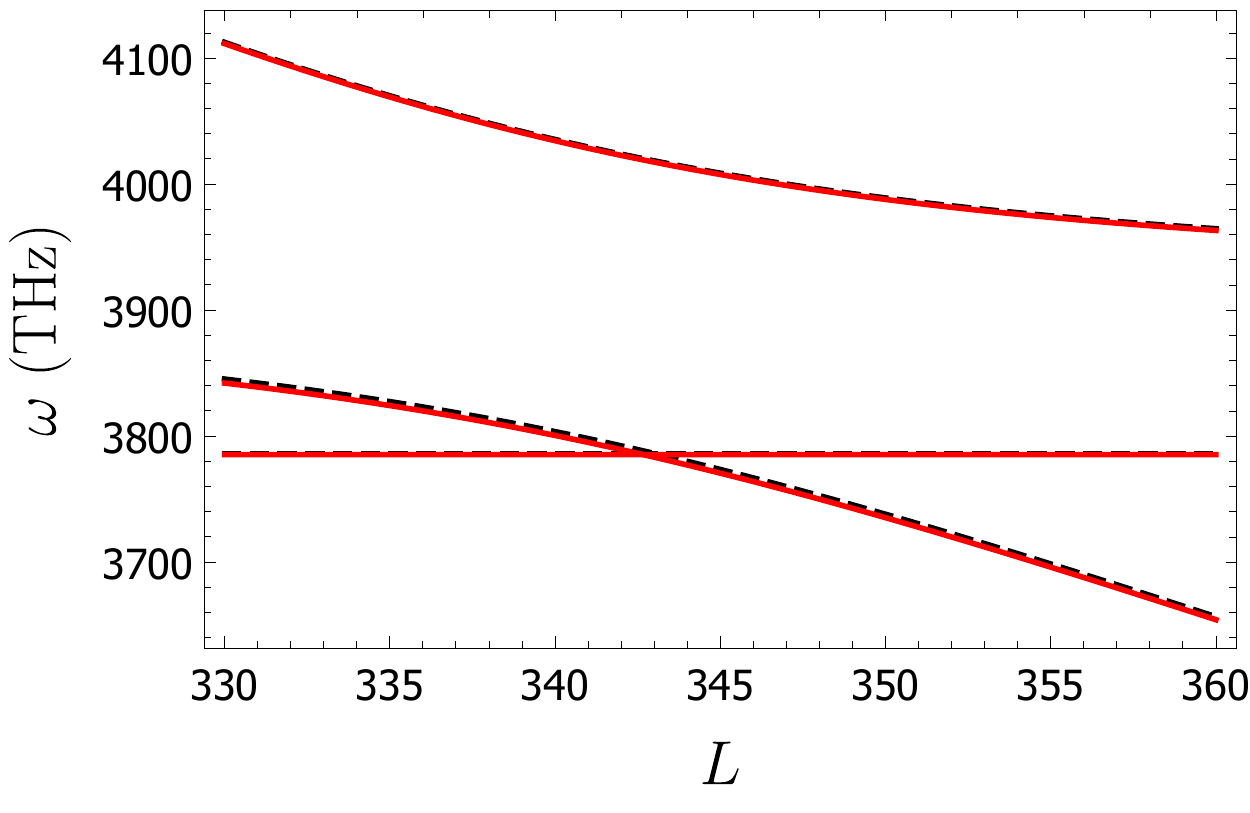}
	\caption[]{\textbf{Corrections to dispersion from non-resonant terms and image dipoles.} Dispersion of the light-matter Hamiltonian for the mode $\ell=(0,1,1)$ from the main text (dashed black lines), and the same Hamiltonian with non-resonant terms and image dipoles included (red lines).}
	\label{fig:non-res_and_image}
\end{figure*}

\noindent Here, the summation is over all $\ell_\alpha$ except when they are all zero. Thus the full non-resonant Hamiltonian including image dipoles is $H=H_\s{dp} + H_\s{im} + H_\s{ph} + H_\s{int}$, where the component Hamiltonians are given in equations \ref{eqn:dipolar_hamiltonian}, \ref{eqn:image_dipoles}, \ref{eqn:photonic_hamiltonian} and \ref{eqn:interaction_hamiltonian} respectively. In figure \ref{fig:non-res_and_image} we plot the dispersion of this Hamiltonian (for the mode $\ell=(0,1,1)$) including non-resonant and image dipoles (red lines) as well as that of the Hamiltonian from the main text which neglects these terms (dashed black lines). We see that the dispersions are near-identical. 

\section{Effects of gain and loss}

To investigate dynamics, as well as the effects of gain and loss, we can take the usual approach of forming the Lindblad master equation for the density matrix

\begin{equation}
    \dot{\rho} = i [\rho,H] + \sum_{m\in \{1,2,c\}} \gamma_m \left( 2 X_m \rho X_m^\dagger - X_m^\dagger X_m \rho - \rho X_m^\dagger X_m \right),
\end{equation}

\noindent where $X_m \in (b_1, b_2, b_c)$ and $\gamma_m$ are the damping rates. We consider the full non-resonant Hamiltonian (neglecting image dipoles)

\begin{equation}
    H = H_\s{dp} + H_\s{ph} + H_\s{int} + H_\s{drive},
\end{equation}

\noindent where the first three terms are given by equations \ref{eqn:dipolar_hamiltonian}, \ref{eqn:fund_ph_ham} and \ref{eqn:fund_int_ham} respectively; and 

\begin{equation}
    H_\s{drive} = \sum_n f_n(t)(b_n + b_n^\dagger),
\end{equation}

\noindent where we consider the symmetric (and anti-symmetric) driving $f_1(t) = \pm f_2(t) = \exp(i \omega_\s{D} t)$. We introduce the (dimensionless) dipole moment $p_n = \langle b_n + b_n^\dagger \rangle$ and conjugate momentum $\pi_n = i \langle b_n - b_n^\dagger \rangle$; and correspondingly define $p_c = \langle c + c^\dagger \rangle$ and $\pi_c = i \langle c - c^\dagger \rangle$. The equations of motion for these expectation values can be obtained using $\langle \dot{O} \rangle = \s{Tr}(\dot{\rho}O)$, and exploiting the cyclic properties of the trace. They are

\begin{eqnarray}\label{eqn:driven_stationary}
\dot{p}_n &=& -\omega_0 \pi_n - 2\xi p_c - \gamma_n p_n, \\
\dot{p}_c &=& -\omega_c \pi_c - \gamma_c p_c, \nonumber \\
\dot{\pi}_n &=& \omega_0 p_n + 2\Omega p_{n+1} - \gamma_n \pi_n + 2f_n, \nonumber\\
\dot{\pi}_c &=& \omega_c p_c -2\xi(\pi_1 + \pi_2) -\gamma_c \pi_c + 2f_c. \nonumber
\end{eqnarray}

\begin{figure*}[t]
	\centering
	\includegraphics[width=0.9\textwidth]{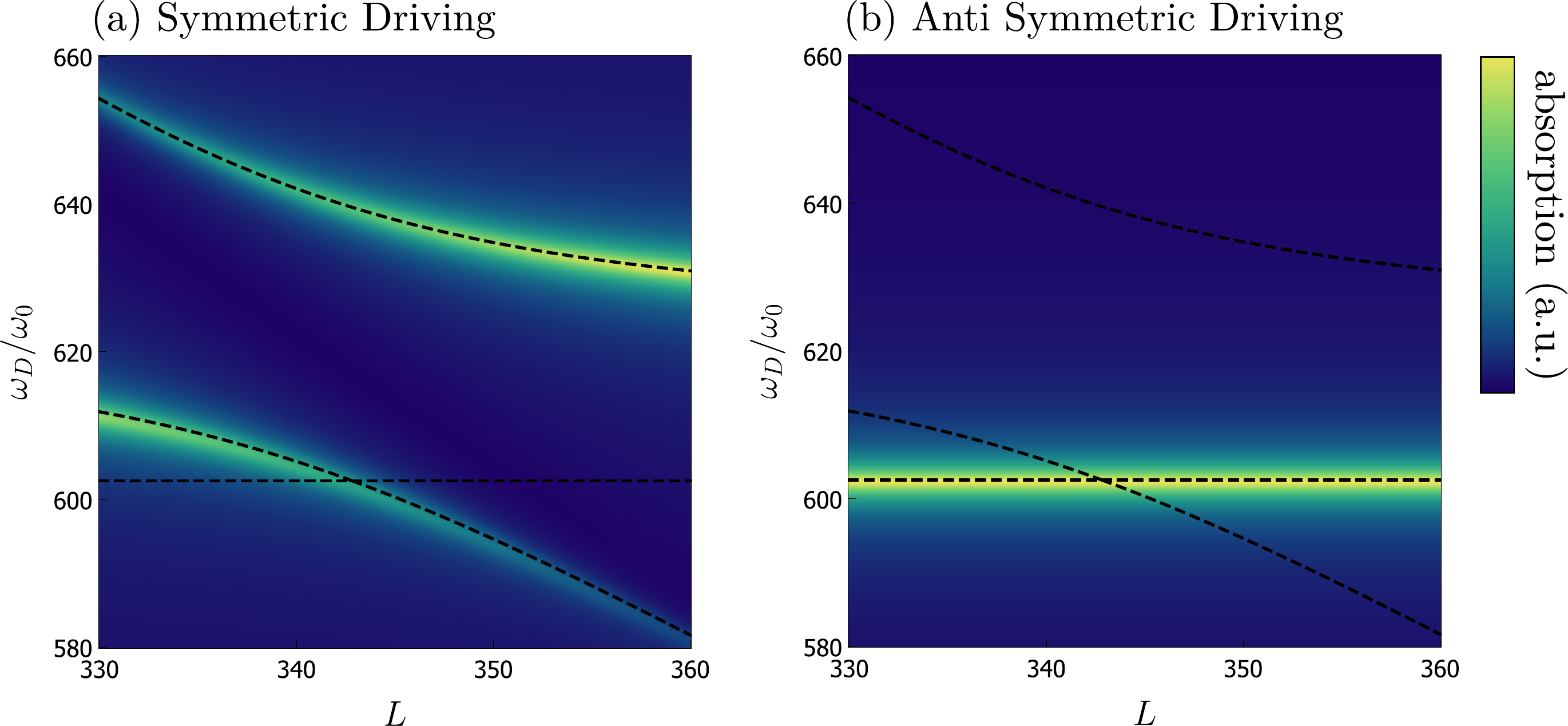}
	\caption[]{\textbf{Absorption in a driven-dissipative dimer.} The dashed lines show the same eigenvalue spectrum of the (Hermitian) Hamiltonian as in Figure 2 of the main text. The colour scale (arbitrary units) corresponds to the total absorption in the meta-atoms, calculated from the stationary solutions of equation \ref{eqn:driven_stationary}, for (a) a symmetric driving, and (b) an anti-symmetric driving.}
	\label{fig:driving}
\end{figure*}

\noindent Using a complex field amplitude representation, we can solve for the steady-state solutions $h_n = A_n \exp(i \omega_D t)$, and hence obtain the total dipole strength, $\mod{A_1} + \mod{A_2}$, as a function of cavity height and driving frequency $\omega_\s{D}$. As a first approximation we consider this dipole strength proportional to the absorbed field at the meta-atoms, which we plot in Figure \ref{fig:driving}. 

\FloatBarrier

\printbibliography

@string{NanoLett = "Nano Lett."}

@string{NatCommun = "Nat. Commun."}

@string{NatMater = "Nat. Mater."}

@string{PhysRev = "Phys. Rev."}

@string{NatPhotonics = "Nat. Photonics"}

@string{PhysRevLett = "Phys. Rev. Lett."}

@string{PhysRevA = "Phys. Rev. A"}

@string{PhysRevB = "Phys. Rev. B"}

@string{ApplPhysLett = "Appl. Phys. Lett."}

@string{Nature = "Nature"}

@string{Science = "Science"}

@string{ACSPhotonics = "ACS Photonics"}

@string{JPhysChemLett = "J. Phys. Chem. Lett."}

@string{ProcNatlAcadSci = "Proc. Natl. Acad. Sci."}

@string{ChemSocRev = "Chem. Soc. Rev."}

@string{JChemPhys = "J. Chem. Phys."}

@string{OptCommun = "Opt. Commun."}

@string{AngewChem = "Angew. Chem."}

@string{Nanoscale = "Nanoscale"}

@string{JPhysB = "J. Phys. B"}

@string{JPhysChemB = "J. Phys. Chem. B"}

@article{Kakazu1994,
	title = {Quantization of electromagnetic fields in cavities and spontaneous emission},
	author = {Kakazu, K. and Kim, Y. S.},
	journal = PhysRevA,
	volume = {50},
	issue = {2},
	pages = {1830--1839},
	year = {1994},
	publisher = {American Physical Society},
	doi = {10.1103/PhysRevA.50.1830},
	url = {https://link.aps.org/doi/10.1103/PhysRevA.50.1830}
}

@article{Brandstetter2015,
	title = {Decay of dark and bright plasmonic modes in a metallic nanoparticle dimer},
	author = {Brandstetter-Kunc, Adam and Weick, Guillaume and Weinmann, Dietmar and Jalabert, Rodolfo A.},
	journal = PhysRevB,
	volume = {91},
	issue = {3},
	pages = {035431},
	year = {2015},
	doi = {10.1103/PhysRevB.91.035431},
	url = {https://link.aps.org/doi/10.1103/PhysRevB.91.035431}
}

@article{Hopfield1958,
	title = {Theory of the Contribution of Excitons to the Complex Dielectric Constant of Crystals},
	author = {Hopfield, J. J.},
	journal = PhysRev,
	volume = {112},
	issue = {5},
	pages = {1555--1567},
	year = {1958},
	publisher = {American Physical Society},
	doi = {10.1103/PhysRev.112.1555},
	url = {https://link.aps.org/doi/10.1103/PhysRev.112.1555}
}

@article{Downing2019,
  title = {Topological Phases of Polaritons in a Cavity Waveguide},
  author = {Downing, C. A. and Sturges, T. J. and Weick, G. and Stobi{\'n}ska, M. and Mart{\'i}n-Moreno, L.},
  journal = PhysRevLett,
  volume = {123},
  issue = {21},
  pages = {217401},
  numpages = {7},
  year = {2019},
  month = {11},
  publisher = {American Physical Society},
  doi = {10.1103/PhysRevLett.123.217401},
  url = {https://link.aps.org/doi/10.1103/PhysRevLett.123.217401}
}

@Article{Mann2018,
	author={Mann, Charlie-Ray
	and Sturges, Thomas J.
	and Weick, Guillaume
	and Barnes, William L.
	and Mariani, Eros},
	title={Manipulating type-I and type-II Dirac polaritons in cavity-embedded honeycomb metasurfaces},
	journal=NatCommun,
	year={2018},
	volume={9},
	number={1},
	pages={2194-2194},
	url={https://doi.org/10.1038/s41467-018-03982-7}
}

@article{Pirmoradian2018,
	title = {Topological magnon modes in a chain of magnetic spheres},
	author = {Pirmoradian, Faezeh and Zare Rameshti, Babak and Miri, MirFaez and Saeidian, Shahpoor},
	journal = PhysRevB,
	volume = {98},
	issue = {22},
	pages = {224409},
	year = {2018},
	doi = {10.1103/PhysRevB.98.224409},
	url = {https://link.aps.org/doi/10.1103/PhysRevB.98.224409}
}

@article{Perczel2017,
	title = {Topological Quantum Optics in Two-Dimensional Atomic Arrays},
	author = {Perczel, J. and Borregaard, J. and Chang, D. E. and Pichler, H. and Yelin, S. F. and Zoller, P. and Lukin, M. D.},
	journal = PhysRevLett,
	volume = {119},
	issue = {2},
	pages = {023603},
	year = {2017},
	doi = {10.1103/PhysRevLett.119.023603},
	url = {https://link.aps.org/doi/10.1103/PhysRevLett.119.023603}
}

@article{Shahnazaryan2019,
	author = {Shahnazaryan, Vanik A. and Saroka, Vasil A. and Shelykh, Ivan A. and Barnes, William L. and Portnoi, Mikhail E.},
	title = {Strong Light-Matter Coupling in Carbon Nanotubes as a Route to Exciton Brightening},
	journal = ACSPhotonics,
	volume = {6},
	number = {4},
	pages = {904-914},
	year = {2019},
	doi = {10.1021/acsphotonics.8b01543},
	URL = {https://doi.org/10.1021/acsphotonics.8b01543}
}

@article{Herrera2016,
  title = {Cavity-Controlled Chemistry in Molecular Ensembles},
  author = {Herrera, Felipe and Spano, Frank C.},
  journal = PhysRevLett,
  volume = {116},
  issue = {23},
  pages = {238301},
  numpages = {6},
  year = {2016},
  month = {6},
  publisher = {American Physical Society},
  doi = {10.1103/PhysRevLett.116.238301},
  url = {https://link.aps.org/doi/10.1103/PhysRevLett.116.238301}
}

@Article{Kowalewski2016,
  author    = {Kowalewski, Markus and Bennett, Kochise and Mukamel, Shaul},
  title     = {Cavity Femtochemistry: Manipulating Nonadiabatic Dynamics at Avoided Crossings},
  journal   = JPhysChemLett,
  year      = {2016},
  volume    = {7},
  number    = {11},
  pages     = {2050--2054},
  month     = jun,
  comment   = {doi: 10.1021/acs.jpclett.6b00864},
  doi       = {10.1021/acs.jpclett.6b00864},
  publisher = {American Chemical Society},
  url       = {https://doi.org/10.1021/acs.jpclett.6b00864},
}

@article{Feist2018,
author = {Feist, Johannes and Galego, Javier and Garcia-Vidal, Francisco J.},
title = {Polaritonic Chemistry with Organic Molecules},
journal = ACSPhotonics,
volume = {5},
number = {1},
pages = {205-216},
year = {2018},
doi = {10.1021/acsphotonics.7b00680},
URL = {https://doi.org/10.1021/acsphotonics.7b0068}
}

@article{Flick2017,
	author = {Flick, Johannes and Ruggenthaler, Michael and Appel, Heiko and Rubio, Angel},
	title = {Atoms and molecules in cavities, from weak to strong coupling in quantum-electrodynamics (QED) chemistry},
	volume = {114},
	number = {12},
	pages = {3026--3034},
	year = {2017},
	doi = {10.1073/pnas.1615509114},
	publisher = {National Academy of Sciences},
	URL = {https://www.pnas.org/content/114/12/3026},
	journal = ProcNatlAcadSci
}

@article{Hertzog2019,
  author = {Hertzog, Manuel and Wang, Mao and Mony, J{\"u}rgen and B{\"o}rjesson, Karl},
  title = {Strong light-matter interactions: a new direction within chemistry},
  journal = ChemSocRev,
  year = {2019},
  volume = {48},
  number = {30662987},
  pages = {937--961},
  month = feb,
  issn = {0306-0012},
  url = {https://www.ncbi.nlm.nih.gov/pmc/articles/PMC6365945/}
}

@article{Felipe2020,
author = {Herrera,Felipe  and Owrutsky,Jeffrey },
title = {Molecular polaritons for controlling chemistry with quantum optics},
journal = JChemPhys,
volume = {152},
number = {10},
pages = {100902},
year = {2020},
doi = {10.1063/1.5136320},
URL = {https://doi.org/10.1063/1.5136320},
eprint = {https://doi.org/10.1063/1.5136320}
}

@Article{Nikolis2019,
  author        = {Nikolis, Vasileios C. and Mischok, Andreas and Siegmund, Bernhard and Kublitski, Jonas and Jia, Xiangkun and Benduhn, Johannes and H{\"o}rmann, Ulrich and Neher, Dieter and Gather, Malte C. and Spoltore, Donato and Vandewal, Koen},
  title         = {Strong light-matter coupling for reduced photon energy losses in organic photovoltaics},
  journal       = NatCommun,
  year          = {2019},
  volume        = {10},
  number        = {1},
  pages         = {3706},
  month         = aug,
  issn          = {2041-1723},
  url       = {https://doi.org/10.1038/s41467-019-11717-5},
}

@article{Downing2017,
  title = {Radiative frequency shifts in nanoplasmonic dimers},
  author = {Downing, Charles A. and Mariani, Eros and Weick, Guillaume},
  journal = PhysRevB,
  volume = {96},
  issue = {15},
  pages = {155421},
  numpages = {9},
  year = {2017},
  month = {10},
  publisher = {American Physical Society},
  doi = {10.1103/PhysRevB.96.155421},
  url = {https://link.aps.org/doi/10.1103/PhysRevB.96.155421}
}

@article{Zuloaga2009,
author = {Zuloaga, Jorge and Prodan, Emil and Nordlander, Peter},
title = {Quantum Description of the Plasmon Resonances of a Nanoparticle Dimer},
journal = NanoLett,
volume = {9},
number = {2},
pages = {887-891},
year = {2009},
doi = {10.1021/nl803811g},
URL = {https://doi.org/10.1021/nl803811g},
eprint = {https://doi.org/10.1021/nl803811g}
}

@article{Bachelier2008,
  title = {Fano Profiles Induced by Near-Field Coupling in Heterogeneous Dimers of Gold and Silver Nanoparticles},
  author = {Bachelier, G. and Russier-Antoine, I. and Benichou, E. and Jonin, C. and Del Fatti, N. and Vall\'ee, F. and Brevet, P.-F.},
  journal = PhysRevLett,
  volume = {101},
  issue = {19},
  pages = {197401},
  numpages = {4},
  year = {2008},
  month = {11},
  publisher = {American Physical Society},
  doi = {10.1103/PhysRevLett.101.197401},
  url = {https://link.aps.org/doi/10.1103/PhysRevLett.101.197401}
}

@article{Dahmen2007,
author = {Dahmen, Christian and Schmidt, Benjamin and von Plessen, Gero},
title = {Radiation Damping in Metal Nanoparticle Pairs},
journal = NanoLett,
volume = {7},
number = {2},
pages = {318-322},
year = {2007},
doi = {10.1021/nl062377u},
URL = {https://doi.org/10.1021/nl062377u},
eprint = {https://doi.org/10.1021/nl062377u}
}

@article{Nordlander2004,
author = {Nordlander, P. and Oubre, C. and Prodan, E. and Li, K. and Stockman, M. I.},
title = {Plasmon Hybridization in Nanoparticle Dimers},
journal = NanoLett,
volume = {4},
number = {5},
pages = {899-903},
year = {2004},
doi = {10.1021/nl049681c},
URL = {https://doi.org/10.1021/nl049681c},
eprint = {https://doi.org/10.1021/nl049681c}
}

@article{Tamaru2002,
author = {Tamaru,Hiroharu  and Kuwata,Hitoshi  and Miyazaki,Hideki T.  and Miyano,Kenjiro },
title = {Resonant light scattering from individual Ag nanoparticles and particle pairs},
journal = ApplPhysLett,
volume = {80},
number = {10},
pages = {1826-1828},
year = {2002},
doi = {10.1063/1.1461072},
URL = {https://doi.org/10.1063/1.1461072},
eprint = {https://doi.org/10.1063/1.1461072}
}

@article{Rechberger2003,
title = "Optical properties of two interacting gold nanoparticles",
journal = OptCommun,
volume = "220",
number = "1",
pages = "137 - 141",
year = "2003",
issn = "0030-4018",
doi = "https://doi.org/10.1016/S0030-4018(03)01357-9",
url = "http://www.sciencedirect.com/science/article/pii/S0030401803013579",
author = "W. Rechberger and A. Hohenau and A. Leitner and J.R. Krenn and B. Lamprecht and F.R. Aussenegg"
}

@article{Danckwerts2007,
  title = {Optical Frequency Mixing at Coupled Gold Nanoparticles},
  author = {Danckwerts, Matthias and Novotny, Lukas},
  journal = PhysRevLett,
  volume = {98},
  issue = {2},
  pages = {026104},
  numpages = {4},
  year = {2007},
  month = {1},
  publisher = {American Physical Society},
  doi = {10.1103/PhysRevLett.98.026104},
  url = {https://link.aps.org/doi/10.1103/PhysRevLett.98.026104}
}

@article{Olk2008,
author = {Olk, Phillip and Renger, Jan and Wenzel, Marc Tobias and Eng, Lukas M.},
title = {Distance Dependent Spectral Tuning of Two Coupled Metal Nanoparticles},
journal = NanoLett,
volume = {8},
number = {4},
pages = {1174-1178},
year = {2008},
doi = {10.1021/nl080044m},
URL = {https://doi.org/10.1021/nl080044m},
eprint = {https://doi.org/10.1021/nl080044m}
}

@article{Zhong2017,
author = {Zhong, Xiaolan and Chervy, Thibault and Zhang, Lei and Thomas, Anoop and George, Jino and Genet, Cyriaque and Hutchison, James A. and Ebbesen, Thomas W.},
title = {Energy Transfer between Spatially Separated Entangled Molecules},
journal = AngewChem,
volume = {56},
number = {31},
pages = {9034-9038},
doi = {10.1002/anie.201703539},
url = {https://onlinelibrary.wiley.com/doi/abs/10.1002/anie.201703539},
eprint = {https://onlinelibrary.wiley.com/doi/pdf/10.1002/anie.201703539},
year = {2017}
}

@Article{Orgiu2015,
  author        = {Orgiu, E. and George, J. and Hutchison, J. A. and Devaux, E. and Dayen, J. F. and Doudin, B. and Stellacci, F. and Genet, C. and Schachenmayer, J. and Genes, C. and Pupillo, G. and Samor{\`i}, P. and Ebbesen, T. W.},
  title         = {Conductivity in organic semiconductors hybridized with the vacuum field},
  journal       = NatMater,
  year          = {2015},
  volume        = {14},
  number        = {11},
  pages         = {1123--1129},
  month         = nov,
  issn          = {1476-4660},
  refid         = {Orgiu2015},
  url           = {https://doi.org/10.1038/nmat4392},
}

@Article{Dovzhenko2018,
author ="Dovzhenko, D. S. and Ryabchuk, S. V. and Rakovich, Yu. P. and Nabiev, I. R.",
title  ="Light-matter interaction in the strong coupling regime: configurations{,} conditions{,} and applications",
journal  = Nanoscale,
year  ="2018",
volume  ="10",
issue  ="8",
pages  ="3589-3605",
publisher  ="The Royal Society of Chemistry",
doi  ="10.1039/C7NR06917K",
url  ="http://dx.doi.org/10.1039/C7NR06917K"
}

@article{Browaeys2016,
	doi = {10.1088/0953-4075/49/15/152001},
	year = 2016,
	month = {6},
	publisher = {{IOP} Publishing},
	volume = {49},
	number = {15},
	pages = {152001},
	author = {Antoine Browaeys and Daniel Barredo and Thierry Lahaye},
	title = {Experimental investigations of dipole-dipole interactions between a few Rydberg atoms},
	journal = JPhysB
}

@article {Leseleuc2019,
	author = {de L{\'e}s{\'e}leuc, Sylvain and Lienhard, Vincent and Scholl, Pascal and Barredo, Daniel and Weber, Sebastian and Lang, Nicolai and B{\"u}chler, Hans Peter and Lahaye, Thierry and Browaeys, Antoine},
	title = {Observation of a symmetry-protected topological phase of interacting bosons with Rydberg atoms},
	volume = {365},
	number = {6455},
	pages = {775--780},
	year = {2019},
	doi = {10.1126/science.aav9105},
	publisher = {American Association for the Advancement of Science},
	issn = {0036-8075},
	URL = {https://science.sciencemag.org/content/365/6455/775},
	eprint = {https://science.sciencemag.org/content/365/6455/775.full.pdf},
	journal = Science
}

@Article{Liu2011,
author ="Liu, Yongmin and Zhang, Xiang",
title  ="Metamaterials: a new frontier of science and technology",
journal  =ChemSocRev,
year  ="2011",
volume  ="40",
issue  ="5",
pages  ="2494-2507",
publisher  ="The Royal Society of Chemistry",
doi  ="10.1039/C0CS00184H",
url  ="http://dx.doi.org/10.1039/C0CS00184H"
}

@article{Kelly2003,
author = {Kelly, K. Lance and Coronado, Eduardo and Zhao, Lin Lin and Schatz, George C.},
title = {The Optical Properties of Metal Nanoparticles: The Influence of Size, Shape, and Dielectric Environment},
journal = JPhysChemB,
volume = {107},
number = {3},
pages = {668-677},
year = {2003},
doi = {10.1021/jp026731y},
URL = {https://doi.org/10.1021/jp026731y},
eprint = {https://doi.org/10.1021/jp026731y}
}

@Article{Du2017,
  author   = {Du, Wei and Wang, Tao and Chu, Hong-Son and Nijhuis, Christian A.},
  title    = {Highly efficient on-chip direct electronic-plasmonic transducers},
  journal  = NatPhotonics,
  year     = {2017},
  volume   = {11},
  number   = {10},
  pages    = {623--627},
  month    = oct,
  issn     = {1749-4893},
  refid    = {Du2017},
  url      = {https://doi.org/10.1038/s41566-017-0003-5},
}

@Article{Li2015,
author ="Li, Ming and Cushing, Scott K. and Wu, Nianqiang",
title  ="Plasmon-enhanced optical sensors: a review",
journal  ="Analyst",
year  ="2015",
volume  ="140",
issue  ="2",
pages  ="386-406",
publisher  ="The Royal Society of Chemistry",
doi  ="10.1039/C4AN01079E",
url  ="http://dx.doi.org/10.1039/C4AN01079E"
}

@article{Yuan2020,
  title = {Enhancement of magnon-magnon entanglement inside a cavity},
  author = {Yuan, H. Y. and Zheng, Shasha and Ficek, Zbigniew and He, Q. Y. and Yung, Man-Hong},
  journal = {Phys. Rev. B},
  volume = {101},
  issue = {1},
  pages = {014419},
  numpages = {10},
  year = {2020},
  month = {1},
  publisher = {American Physical Society},
  doi = {10.1103/PhysRevB.101.014419},
  url = {https://link.aps.org/doi/10.1103/PhysRevB.101.014419}
}

@Article{Baranov2020,
  author   = {Baranov, Denis G. and Munkhbat, Battulga and Zhukova, Elena and Bisht, Ankit and Canales, Adriana and Rousseaux, Benjamin and Johansson, Göran and Antosiewicz, Tomasz J. and Shegai, Timur},
  title    = {Ultrastrong coupling between nanoparticle plasmons and cavity photons at ambient conditions},
  journal  = {Nature Communications},
  year     = {2020},
  volume   = {11},
  number   = {1},
  pages    = {2715},
  month    = jun,
  issn     = {2041-1723},
  refid    = {Baranov2020},
  url      = {https://doi.org/10.1038/s41467-020-16524-x},
}

@article {Pendry2006,
	author = {Pendry, J. B. and Schurig, D. and Smith, D. R.},
	title = {Controlling Electromagnetic Fields},
	volume = {312},
	number = {5781},
	pages = {1780--1782},
	year = {2006},
	doi = {10.1126/science.1125907},
	publisher = {American Association for the Advancement of Science},
	issn = {0036-8075},
	URL = {https://science.sciencemag.org/content/312/5781/1780},
	eprint = {https://science.sciencemag.org/content/312/5781/1780.full.pdf},
	journal = {Science}
}

@article{Fruhnert2015,
author = {Fruhnert, Martin and Kretschmer, Florian and Geiss, Reinhard and Perevyazko, Igor and Cialla-May, Dana and Steinert, Michael and Janunts, Norik and Sivun, Dmitry and Hoeppener, Stephanie and Hager, Martin D. and Pertsch, Thomas and Schubert, Ulrich S. and Rockstuhl, Carsten},
title = {Synthesis, Separation, and Hypermethod Characterization of Gold Nanoparticle Dimers Connected by a Rigid Rod Linker},
journal = {The Journal of Physical Chemistry C},
volume = {119},
number = {31},
pages = {17809-17817},
year = {2015},
doi = {10.1021/acs.jpcc.5b04346},

URL = { 
        https://doi.org/10.1021/acs.jpcc.5b04346},
eprint = { 
        https://doi.org/10.1021/acs.jpcc.5b04346}
}

@article{Cunningham2011,
author = {Cunningham, Alastair and Mühlig, Stefan and Rockstuhl, Carsten and Bürgi, Thomas},
title = {Coupling of Plasmon Resonances in Tunable Layered Arrays of Gold Nanoparticles},
journal = {The Journal of Physical Chemistry C},
volume = {115},
number = {18},
pages = {8955-8960},
year = {2011},
doi = {10.1021/jp2011364},

URL = { 
        https://doi.org/10.1021/jp2011364
    
},
eprint = { 
        https://doi.org/10.1021/jp2011364
    
}

}

@article{Cunningham2012,
author = {Cunningham, Alastair and Mühlig, Stefan and Rockstuhl, Carsten and Bürgi, Thomas},
title = {Exciting Bright and Dark Eigenmodes in Strongly Coupled Asymmetric Metallic Nanoparticle Arrays},
journal = {The Journal of Physical Chemistry C},
volume = {116},
number = {33},
pages = {17746-17752},
year = {2012},
doi = {10.1021/jp301764d},

URL = { 
        https://doi.org/10.1021/jp301764d
    
},
eprint = { 
        https://doi.org/10.1021/jp301764d
    
}

}

\end{document}